# A Bayesian Network Model of the Bit Error Rate for Cognitive Radio Networks

Hector Reyes, Sriram Subramaniam, and Naima Kaabouch
Electrical Engineering, University of North Dakota
Grand Forks, North Dakota, United States

*Abstract—* In addition to serve as platforms for dynamic spectrum access, cognitive radios can also serve as a method for improving the performance of wireless communication systems by smartly adjusting their operating parameters according to the environment and requirements. The uncertainty always present in the environment makes the practical implementation of the latter application difficult. In this paper, we propose a probabilistic graphical model, Bayesian network that captures the causal relationships among the variables bit energy to noise spectral density ratio (EbN0), carrier to interference ratio (C/I), modulation scheme (MOD), Doppler phase shift (Dop_Phi), and bit error rate (BER). BER indicates how the communication link is performing. The goal of our proposed Bayesian network is to use the BER as evidence in order to infer the behavior of the other variables, so the cognitive radio can learn how the conditions of the environment are, and based on that knowledge make better informed decisions. This model along with the method used to build it are described in this paper.

## I. INTRODUCTION

Cognitive radios (CR) have been conceived as wireless communication devices that observe the radio scene, make decisions, and perform actions to adapt to it [1, 2]. Although the original purpose of cognitive radio technology was to provide a means for implementing dynamic spectrum access, it can also be applied to improving the performance of communication systems, by setting up smartly configuration parameters such as transmission power, bandwidth, etc. [3-6]. In the case of the latter application, the CR needs to observe some variables indicating how the system is performing. One of the most prevalent performance indicators in wireless systems is the bit error rate (BER). BER is defined as the number of erroneous bits divided by the total number of transmitted bits. This parameter is affected by several random factors such as bit energy to noise spectral density ratio, EbN0, carrier to interference ratio, C/I, and Doppler phase shift. Because of this randomness, the interaction among these variables should be analyzed with a probabilistic method. In this paper, we propose using a Bayesian network approach [7-10] to model the interaction among the variables that affect the BER. Such a model will allow to infer the probability distribution of selected variables (causes or hidden variables) based on the value of BER, the evidence. The knowledge of these probability distributions can provide information the CR could use to make better informed decisions.

## II. METHODOLOGY

To conceive the model, we started by analyzing which variables affect the BER. In this step, we have applied our knowledge about communication systems. We came up with the directed graph shown in Fig. 1. This graph constitutes the structure or qualitative part of the Bayesian model. The qualitative part indicates how the variables influence one to another. In this figure, we can distinguish two types of variables: parent and child. One or several parent variables affect one or several child variables. In this case, bit energy to noise spectral density ratio, EbN0, carrier to interference ratio, C/I, modulation scheme, MOD, and the Doppler phase shift, Dop_Phi, are parents of the variable BER. We have assumed narrow or flat channel, so we can ignore the effect of the delay spread.

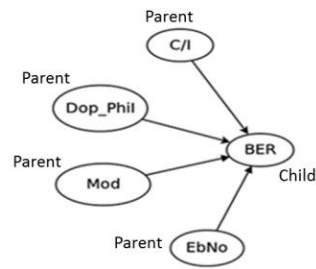

Fig. 1. Structure of Bayesian Model for BER.

The other important component of the model is the quantitative part represented by conditional probability distributions (CPD), which numerically shows how the parent variables affect the probability distribution of the child variable. The CPD of a variable indicates the probability of the variable falling at each of its states as a result of a particular combination of the states of its parents. The proposed model uses discrete variables; therefore, the scope of each variable is divided in non-overlapping intervals represented by states taken by the variable; the modeler decides the number of intervals for each variable. Table I shows how we have defined the states and corresponding intervals for the variables EbN0, Dop_Phi, C/I, and BER. To obtain the CPD of BER we have resorted to combining our previous knowledge, represented by Fig. 1, with data acquired through simulations. For simplicity, we simulated three different kinds of modulation: differential binary phase shift keying (DBPSK), differential quaternary phase shift keying (DQPSK), and differential 8 phase shift keying (D8PSK). However, the method is applicable to other modulation schemes. Fig. 2 shows the simulation of the effect of EbN0, C/I, and Dop_Phi on BER. In ① a random symbol is selected and represented as an exponential number with phase set according to the modulation scheme: tables labeled as DBPSK, DQPSK, and D8PSK. In ② this number is multiplied by another exponential number with phase equal to the Doppler phase shift. In ③ co-channel interference and noise are added. The received symbols (represented as exponential numbers) are demodulated, stage ④, compared with the transmitted bits, stage ⑤, finally the errors are accumulated to calculate the BER, stage ⑥. We performed this process multiple times with different combinations of the states of EbN0, C/I and Dop_Phi. Fig. 3 illustrates how after running this process, represented as ①, the data yielded by it passes through a discretizer block, ②, and then through a CPD learner block, ③. The function of the discretizer is to replace the numeric samples by the name of the interval, state, wherein they fall. The reason for doing this process is that the CPDs are represented in terms of the states of the variables, not in terms of numbers. The CPD learner takes the data and the structure, Bayesian network, which governs the interaction among the variables BER, EbN0, C/I, Dop_Phi and MOD, and applies the maximum likelihood estimation (MLE) algorithm to yield the conditional probability table as shown at the bottom of Fig. 3.

TABLE I
DEFINITION OF STATES AND INTERVALS FOR THE VARIABLES EbN0, C/I, DOP_PHI, BER.

| Variable | State | Interval |
|---|---|---|
| EbN0 (dB) | EbN0_1 | -72.8 dB to 0 dB |
| | EbN0_2 | 0 dB to 10 dB |
| | EbN0_3 | 10 dB to 13 dB |
| | EbN0_4 | 13 dB to 16 dB |
| | EbN0_5 | 16 dB to 19 dB |
| | EbN0_6 | 19 dB to 109.1 dB |
| C/I (dB) | C/I_1 | -159 dB to 0 dB |
| | C/I_2 | 0 dB to 20 dB |
| | C/I_3 | 20 dB to 30 dB |
| | C/I_4 | 30 dB to 40 dB |
| | C/I_5 | 40 dB to 50 dB |
| | C/I_6 | 50 dB to 159 dB |
| Dop_Phi (radians) | Phi_1 | 0 rad. to 0.05 rad. |
| | Phi_2 | 0.05 rad. to 0.1 rad. |
| | Phi_3 | 0.1 rad. to 0.136 rad. |
| BER | BER_1 | 0 to $10^{-5}$ |
| | BER_2 | $10^{-5}$ to $10^{-3}$ |
| | BER_3 | $10^{-3}$ to $10^{-2}$ |
| | BER_4 | $10^{-2}$ to $10^{-1}$ |
| | BER_5 | $10^{-1}$ to 1 |

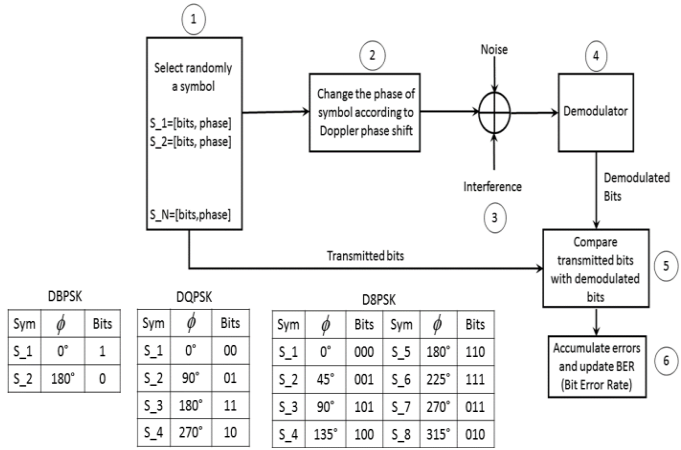

Fig. 2. Simulating the effect of EbN0, C/I and Dop_Phi on BER

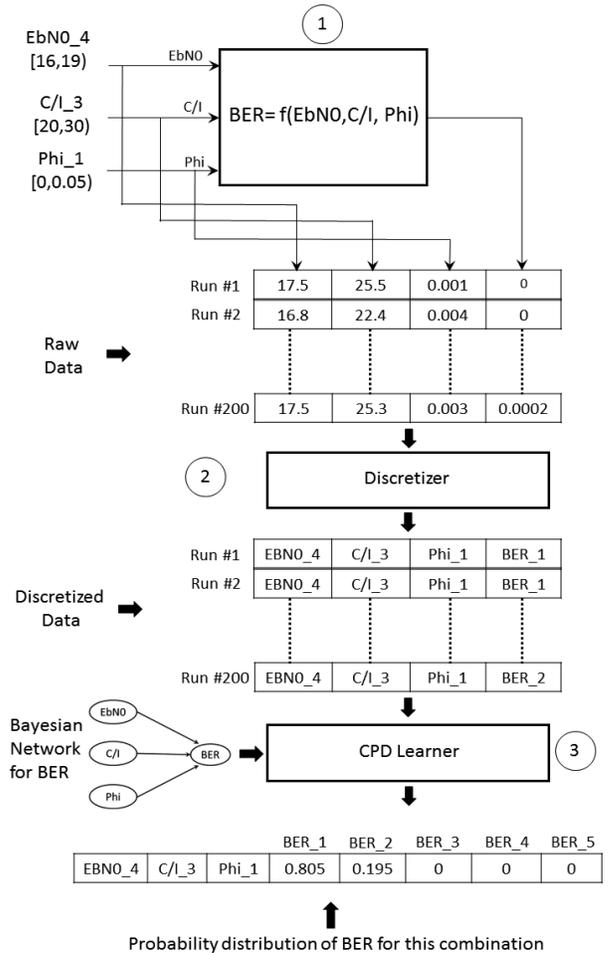

Fig. 3. Process for learning the conditional probability distribution (CPD) for BER. In this example, the modulation system is DQPSK.

III. RESULTS AND DISCUSSION

Tables II through VI show the conditional probability distributions (CPD) for BER obtained after performing the simulation for different modulation schemes. A CPD describes how the probability distribution of a random variable changes according to its influencing variables. When a random variable depends on discrete variables, its CPD must map each combination of the states of these variables, its parents, to its probability distribution. The CPD of the bit error rate, BER has 108 combinations due to the different states of the parents: 6 states for EbN0 times 6 states for C/I times 3 states for Dop_Phi. The probability distribution of BER falls over 5 states: BER_1, BER_2, BER_3, BER_4, and BER_5. BER_1 is the preferable state for communication systems, since it corresponds to the lowest error rate. Therefore, the parameters of the system must be manipulated so that the probability of the BER being low (states BER_1, or BER_2) increases and the probability of being high (states BER_3 to BER_5) decreases.

Since these tables include a high number of combinations, we will analyze only some of the more interesting combinations. Let us start with the combinations where the probability of BER being high is the greatest. This situation happens when either EbN0 or C/I is low and gets more critical as the number of bits per symbol increases. When EbN0=EbN0_1, no matter what the states of C/I and Dop_Phi are, BER= BER_5 with probability 1. This outcome was obtained for all the modulation schemes used in this work: DBPSK, DQPSK, and D8PSK. We can see that in table II, the probability distribution of BER is [0, 0, 0, 0, 1], which corresponds to the probabilities of this variable being in the states BER_1, BER_2, BER_3, BER_4, and BER_5, respectively, in other words: [p(BER=BER_1), p(BER=BER_2), p(BER=BER_3), p(BER=BER_4), p(BER=BER_5)]. When C/I=C/I_1, the probability distribution of BER is also [0, 0, 0, 0, 1] as shown in Table III.

TABLE II
PROBABILITY DISTRIBUTION OF BER FOR DBPSK, DQPSK, AND D8PSK WHEN EbN0 IS EbN0_1

| States of the Parents | | | Probability distribution of BER | | | | |
|---|---|---|---|---|---|---|---|
| EbN0_1 to EbN0_6 | C/I_1 | Phi_1 to Phi_3 | 0 | 0 | 0 | 0 | 1 |

TABLE III
PROBABILITY DISTRIBUTION OF BER FOR DBPSK, DQPSK, AND D8PSK WHEN C/I IS C/I_1

| States of the Parents | | | Probability distribution of BER | | | | |
|---|---|---|---|---|---|---|---|
| EbN0_1 | C/I_1 to C/I_6 | Phi_1 to Phi_3 | 0 | 0 | 0 | 0 | 1 |

Table IV shows that when the modulator is DBPSK and the EbN0 reaches its maximum, EbN0=EbN0_6, p(BER=BER_1) gets its highest values and even reaches 1; we can observe how the probability distribution for BER changes between the row I and the row II of table IV. It is important to mention that if the variable C/I is at a low value, it also affects the probability distribution of BER. These results are consistent with our previous knowledge about the interaction among the aforementioned variables: the higher EbN0 and C/I, the lower BER.

TABLE IV
PROBABILITY DISTRIBUTION OF BER FOR DBPSK WHEN EbN0 IS EbN0_6

| States of the parents | | | Probability distribution of BER | | | | | |
|---|---|---|---|---|---|---|---|---|
| EbN0_6 | C/I_2 | Phi_1 | 0.65 | 0 | 0.025 | 0.07 | 0.255 | I |
| EbN0_6 | C/I_2 | Phi_2 | 0.63 | 0.03 | 0.015 | 0.08 | 0.245 | |
| EbN0_6 | C/I_2 | Phi_3 | 0.605 | 0.04 | 0.025 | 0.07 | 0.26 | |
| EbN0_6 | C/I_3 | Phi_1 | 1 | 0 | 0 | 0 | 0 | |
| EbN0_6 | C/I_3 | Phi_2 | 1 | 0 | 0 | 0 | 0 | |
| EbN0_6 | C/I_3 | Phi_3 | 1 | 0 | 0 | 0 | 0 | |
| EbN0_6 | C/I_4 | Phi_1 | 1 | 0 | 0 | 0 | 0 | |
| EbN0_6 | C/I_4 | Phi_2 | 1 | 0 | 0 | 0 | 0 | |
| EbN0_6 | C/I_4 | Phi_3 | 1 | 0 | 0 | 0 | 0 | |
| EbN0_6 | C/I_5 | Phi_1 | 1 | 0 | 0 | 0 | 0 | |
| EbN0_6 | C/I_5 | Phi_2 | 1 | 0 | 0 | 0 | 0 | |
| EbN0_6 | C/I_5 | Phi_3 | 1 | 0 | 0 | 0 | 0 | |
| EbN0_6 | C/I_6 | Phi_1 | 1 | 0 | 0 | 0 | 0 | II |
| EbN0_6 | C/I_6 | Phi_2 | 1 | 0 | 0 | 0 | 0 | |
| EbN0_6 | C/I_6 | Phi_3 | 1 | 0 | 0 | 0 | 0 | |

TABLE V
PROBABILITY DISTRIBUTION OF BER FOR DQPSK WHEN EbN0 IS EbN0_6

| States of the parents | | | Probability distribution of BER | | | | | |
|---|---|---|---|---|---|---|---|---|
| EbN0_6 | C/I_2 | Phi_1 | 0.375 | 0.03 | 0.015 | 0.135 | 0.445 | I |
| EbN0_6 | C/I_2 | Phi_2 | 0.355 | 0.06 | 0.07 | 0.12 | 0.395 | |
| EbN0_6 | C/I_2 | Phi_3 | 0.25 | 0.07 | 0.1 | 0.165 | 0.415 | |
| EbN0_6 | C/I_3 | Phi_1 | 1 | 0 | 0 | 0 | 0 | |
| EbN0_6 | C/I_3 | Phi_2 | 1 | 0 | 0 | 0 | 0 | |
| EbN0_6 | C/I_3 | Phi_3 | 0.995 | 0.005 | 0 | 0 | 0 | |
| EbN0_6 | C/I_4 | Phi_1 | 1 | 0 | 0 | 0 | 0 | |
| EbN0_6 | C/I_4 | Phi_2 | 1 | 0 | 0 | 0 | 0 | |
| EbN0_6 | C/I_4 | Phi_3 | 1 | 0 | 0 | 0 | 0 | |
| EbN0_6 | C/I_5 | Phi_1 | 1 | 0 | 0 | 0 | 0 | |
| EbN0_6 | C/I_5 | Phi_2 | 1 | 0 | 0 | 0 | 0 | |
| EbN0_6 | C/I_5 | Phi_3 | 0.995 | 0.005 | 0 | 0 | 0 | |
| EbN0_6 | C/I_6 | Phi_1 | 1 | 0 | 0 | 0 | 0 | II |
| EbN0_6 | C/I_6 | Phi_2 | 1 | 0 | 0 | 0 | 0 | |
| EbN0_6 | C/I_6 | Phi_3 | 1 | 0 | 0 | 0 | 0 | |

TABLE VI
PROBABILITY DISTRIBUTION OF BER FOR D8PSK WHEN EbN0 IS EbN0_6

| States of the parents | | | Probability distribution of BER | | | | | |
|---|---|---|---|---|---|---|---|---|
| EbN0_6 | C/I_2 | Phi_1 | 0.025 | 0.015 | 0.035 | 0.25 | 0.675 | I |
| EbN0_6 | C/I_2 | Phi_2 | 0 | 0.01 | 0.07 | 0.235 | 0.685 | |
| EbN0_6 | C/I_2 | Phi_3 | 0 | 0 | 0.025 | 0.355 | 0.62 | |
| EbN0_6 | C/I_3 | Phi_1 | 0.96 | 0.035 | 0.005 | 0 | 0 | |
| EbN0_6 | C/I_3 | Phi_2 | 0.725 | 0.225 | 0.05 | 0 | 0 | |
| EbN0_6 | C/I_3 | Phi_3 | 0.36 | 0.34 | 0.295 | 0.005 | 0 | |
| EbN0_6 | C/I_4 | Phi_1 | 0.96 | 0.04 | 0 | 0 | 0 | |
| EbN0_6 | C/I_4 | Phi_2 | 0.945 | 0.055 | 0 | 0 | 0 | |
| EbN0_6 | C/I_4 | Phi_3 | 0.935 | 0.06 | 0.005 | 0 | 0 | |
| EbN0_6 | C/I_5 | Phi_1 | 0.985 | 0.015 | 0 | 0 | 0 | |
| EbN0_6 | C/I_5 | Phi_2 | 0.96 | 0.04 | 0 | 0 | 0 | |
| EbN0_6 | C/I_5 | Phi_3 | 0.915 | 0.06 | 0.025 | 0 | 0 | |
| EbN0_6 | C/I_6 | Phi_1 | 0.99 | 0.01 | 0 | 0 | 0 | II |
| EbN0_6 | C/I_6 | Phi_2 | 0.945 | 0.05 | 0.005 | 0 | 0 | |
| EbN0_6 | C/I_6 | Phi_3 | 0.955 | 0.045 | 0 | 0 | 0 | |

Tables V and VI illustrate that a similar behavior takes place with DQPSK and D8PSK modulators. However, these tables indicate that at the same levels of EbN0 and C/I the probability of getting low BER, p(BER=BER_1) for DQPSK and and D8PSK modulators is smaller than the one for the DBPSK modulator, whereas the probability of getting high BER, p(BER=BER_5) is bigger. This fact is consistent with what we already know about the BER of the DBPSK, DQPSK, and D8PSK modulation schemes: the more bits per symbol, the higher the probability of error for the same operating conditions, such as levels of noise and interference. The reason is that as the number of constellation points of a modulation method increases, its tolerance to noise and interference decreases, which generates more errors. The DQPSK and D8PSK modulators use 2 and 3 bits per symbol respectively; therefore, their BERs are higher than the BER of the DBPSK modulator, which uses only 1 bit per symbol. D8PSK tends to have higher BER than DQPSK, since it uses 3 bits per symbol instead of 2; hence, p(BER=BER_1) and p(BER=BER_5) for D8PSK are respectively lower and higher than those of DQPSK.

The importance of these results is that they confirm and express our knowledge in terms of probabilities, so that we can use it in Bayesian and probabilistic graphical models, PGM.

The structure shown in Fig. 1 along with the CPDs summarized by tables II through VI form the Bayesian network (BN) for the variable BER. A cognitive radio (CR) could use this BN as a probabilistic reasoning engine to infer the CPD of variables of interest (EbN0, C/I, and Dop_Phi) based on its measurements of BER. With this knowledge the CR can make better informed decisions as of how to set up its parameters to increase the probability that BER will be at its lowest value, BER_1. We can extend the BN in Fig. 1 to include other variables that influence EbN0, C/I, and Dop_Phi. This augmented model could provide a means for the CR to learn about factors of the surrounding environment that affect its operation.

IV. CONCLUSION

In this paper, we have proposed a Bayesian network that represents the interaction between the variables EbN0, C/I, Dop_Phi, and BER. At the best of our knowledge, nowhere in the literature there is a model that captures how the variables EbN0, C/I, and Dop_Phi simultaneously affect BER. In this paper we have proposed such a model in a probabilistic fashion represented by means of a Bayesian Network. This BN can serve as a reasoning engine for a cognitive radio, CR. As the CR measures BER, it can infer the CPDs of the variables of interest to learn from the environment and decide on its parameters. Future work includes adding more variables and complementing the BN with a utility function (UF). Adding a UF to a BN turns the latter into an influence diagram, a probabilistic graphical model that can assist the CR in making decisions in the midst of uncertainty.

ACKNOWLEDGMENT

The authors acknowledge the support of NSF, grant # 1443861, and EPSCoR/NSF grant # EPS-0184442.